\documentclass[aps,prd,reprint,superscriptaddress,nofootinbib]{revtex4-2}
\usepackage{graphicx}
\usepackage{amsmath}
\usepackage{booktabs}
\usepackage[colorlinks=true,allcolors=blue]{hyperref}
\usepackage{xurl}
\begin{document}

\title{Is the arXiv Apocalypse upon us? A break in preprint growth coincident with agentic AI}

\author{Carlo R. Contaldi}
\email{c.contaldi@imperial.ac.uk}
\affiliation{Department of Physics, Imperial College London, London SW7 2AZ, United Kingdom}

\date{\today}

\begin{abstract}
As an experiment in agentic assisted analysis, we used an agentic AI model to analyse the effect of agentic AI on arXiv submissions; data acquisition, analysis and the first draft were carried out by the model under the author's direction. Monthly arXiv submissions have grown exponentially for three decades, at about $9\%$ per year. Fitting a piecewise log-linear trend with seasonal modulation to the full record, we find two post-pandemic breaks in growth rate. The first, in mid-2022, precedes ChatGPT and is carried by the expansion of AI research within computer science (CS); outside CS, growth from 2022 to 2025 remained at $8\%$ per year. The second, between December 2025 and February 2026, follows the release of models capable of extended autonomous work and is absent from the AI categories themselves. Since December 2025 submissions without a CS listing have run $17\%$ above trend, and in September 2026 total submissions reached a record $40\,363$. The excess is largest in mathematics, where submissions doubled within a year, and follows a gradient from formal and theoretical fields to experimental ones that is not ordered by surveyed rates of AI-tool adoption. In mathematics the increase is distributed across subfields and authors, two-thirds of it arising from a larger number of distinct authors rather than from a few prolific submitters. Across fields the clearest signature is a shift to single authorship: solo-authored papers rose by $70\%$ within a year in mathematics and condensed matter, and by $44\%$ in quantitative biology even as its total output fell, but not in experimental particle physics. The analysis is deliberately agnostic of LLM-detection methods: it uses only submission counts and metadata, with treatment timing fixed by calendar dates, and is therefore free of the outcome-triggered timing bias of author-level studies.
\end{abstract}


\maketitle

\section{Introduction}

The arXiv preprint server has been the primary dissemination channel for physics, mathematics and, increasingly, computer science since 1991~\cite{Ginsparg2011}. Its monthly submission count is among the longest and most homogeneous records of scientific output available. Over thirty-five years the series has grown by four orders of magnitude, following to good approximation an exponential trend on which a reproducible seasonal pattern is superposed.

Since late 2022 large language models (LLMs) have become widely used in the preparation of scientific manuscripts. Estimates based on word-frequency distributions indicate that the fraction of LLM-modified text in arXiv abstracts rose most rapidly in computer science and most slowly in mathematics~\cite{Liang2024}. At the level of individual authors, Kusumegi \textit{et al.}~\cite{Kusumegi2025} reported that scientists who adopt LLMs for drafting increase their preprint output by between $24\%$ and $89\%$, depending on field and background. That result has been challenged on methodological grounds: adoption is dated by the first month in which a submitted abstract crosses a detection threshold, a rule that is mechanically more likely to trigger in high-output months and that reproduces a positive post-treatment effect under placebo assignments~\cite{Renault2026}.

During 2025 the character of the available tools changed. Models released in the second half of that year were designed to carry out extended, multi-step tasks with limited supervision, including writing and executing analysis code and drafting documents. Usage of such tools grew rapidly in the first half of 2026~\cite{Johnston2026}, including among academic scientists~\cite{RepublicOfScience2026}. In September 2026 arXiv recorded $40\,363$ submissions, the largest monthly total in its history, and from 1 October 2026 introduced a limit of two new submissions per submitter per calendar month, citing an increase in thin, fragmented and machine-written papers~\cite{arXivCap2026}.

In this paper we ask whether the aggregate submission record contains a distinct signature of this transition. The analysis is intentionally agnostic of LLM-detection tools. Existing estimates of LLM use rely on classifiers or word-frequency signatures whose calibration is uncertain~\cite{Liang2023Detectors,WeberWulff2023,Kobak2025}, which drift as models change and as authors adapt their usage~\cite{GengTrotta2025,Krishna2023}, and which can be expected to be least reliable for agentic workflows, in which a model writes analysis code and generates results as well as prose. We use none of them. Every quantity in this paper is derived from submission counts and the metadata attached to each submission. Our approach also differs from author-level studies in two further respects. First, we locate changes in the growth rate directly from the data, without imposing a treatment date, and only afterwards compare their location with the release history of agentic models. Second, where we attribute an effect, the comparison is between fields over a common calendar window, so that the timing of the putative treatment is exogenous to the output of any individual author.

\section{Data}

The aggregate series consists of monthly new-submission counts from July 1991 to October 2026, as published by arXiv. The October 2026 value is excluded throughout because it is incomplete and because it is the first month subject to the per-submitter limit; it is discussed separately in Sec.~\ref{sec:cap}.

Category-level counts were obtained from the arXiv API for each month from January 2022 to October 2026 by querying on submission date and category.\footnote{The arXiv category taxonomy is available at \url{https://arxiv.org/category_taxonomy}.} The API matches a paper to every category in which it is listed, so counts include cross-lists and groups overlap. To construct control groups that are not contaminated by the growth of machine-learning research itself, each non-computer-science group was additionally queried with all papers carrying any \texttt{cs.*} listing removed. Three categories are shared between archives under alias names: numerical analysis (\texttt{math.NA}, alias \texttt{cs.NA}), systems and control (\texttt{eess.SY}, alias \texttt{cs.SY}) and information theory (\texttt{cs.IT}, alias \texttt{math.IT}). The API indexes each such paper under the first name only. Papers in numerical analysis and in systems and control therefore remain in the no-CS-listing groups unless they carry an additional \texttt{cs.*} listing, as about a fifth of numerical-analysis papers do, whereas information-theory papers are counted as computer science and do not enter the mathematics group. The groups used are listed in Table~\ref{tab:fields}. Totals returned by the API agree with the published aggregate counts to within $\sim\!2\%$, the residual difference arising from withdrawals and reclassification.

For two months, September 2025 and September 2026, the full metadata of every mathematics submission without a computer-science listing was retrieved ($3\,945$ and $8\,203$ records respectively), comprising author lists, primary category and submission date. These are used for the author-concentration analysis of Sec.~\ref{sec:authors}.

\begin{table}
\caption{Category groups, defined by arXiv API queries, with counts for September 2026. All groups listed under ``No CS listing'' exclude every paper carrying any \texttt{cs.*} listing. The ``No CS listing'' group itself is the aggregate total minus the CS group and is exact.}
\label{tab:fields}
\begin{ruledtabular}
\footnotesize
\begin{tabular}{@{}lll@{}}
Group & Query & Sep 2026 \\
\hline
All & all submissions & 39\,267 \\ 
Computer science & \texttt{cs.*} & 19\,618 \\ 
\quad cs.AI / cs.LG / cs.CL & \texttt{cs.AI}, \texttt{cs.LG}, \texttt{cs.CL} & 6\,008 / 5\,730 / 2\,880 \\
No CS listing & all $-$ \texttt{cs.*} & 19\,649 \\ 
\quad Mathematics & \texttt{math.*} & 8\,203 \\ 
\quad Statistics & \texttt{stat.*} & 680 \\ 
\quad Quantum physics & \texttt{quant-ph} & 1\,924 \\ 
\quad Economics & \texttt{econ.*}, \texttt{q-fin.*} & 418 \\ 
\quad HEP, GR, nuclear & \texttt{hep-*}, \texttt{gr-qc}, \texttt{nucl-*} & 2\,629 \\ 
\quad Condensed matter & \texttt{cond-mat.*} & 2\,623 \\ 
\quad Astrophysics & \texttt{astro-ph.*} & 2\,321 \\ 
\quad Other physics & \texttt{physics.*} & 2\,426 \\ 
\quad EESS & \texttt{eess.*} & 1\,086 \\ 
\quad Quant.\ biology & \texttt{q-bio.*} & 249 \\ 
\end{tabular}
\end{ruledtabular}
\end{table}

\begin{figure}[t]
\includegraphics[width=\columnwidth]{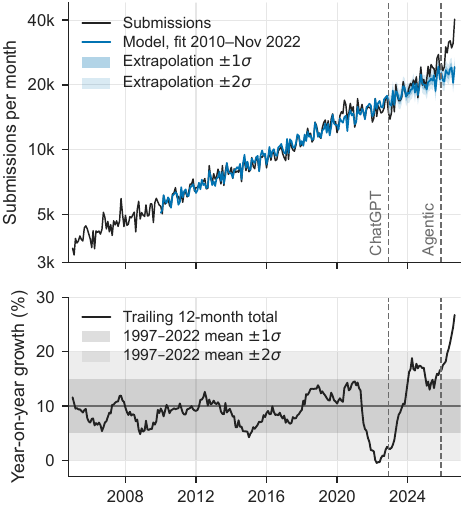}
\caption{Top: monthly arXiv submissions (black) and the model of Eq.~\eqref{eq:model} without knots, fitted to January 2010 to November 2022 (blue), with its $\pm 1\sigma$ and $\pm 2\sigma$ prediction bands after the fitting interval. Bottom: year-on-year growth of the trailing twelve-month total, compared with the mean and $\pm 1\sigma$ and $\pm 2\sigma$ ranges of annual growth over 1997--2022. The dashed lines mark the public release of ChatGPT and the cluster of agentic model releases in late 2025.}
\label{fig:aggregate}
\end{figure}

\section{Method}

\subsection{Trend and seasonal model}

We model the logarithm of the monthly count $N(t)$ as
\begin{align}
\ln N(t) ={}& a + b\,(t - t_0) + \sum_{j} c_j \,(t - \tau_j)\,\Theta(t-\tau_j) \nonumber\\
&+ \sum_{k=2}^{12} s_k\, \delta_{m(t),k} + w \ln\frac{W(t)}{\bar W} + \epsilon(t),
\label{eq:model}
\end{align}
where $t$ is time in years, $\Theta$ is the Heaviside function, $\tau_j$ are knot locations at which the growth rate changes by $c_j$, $s_k$ are seasonal offsets for calendar month $k$ relative to January, $W(t)$ is the number of weekdays in the month, $\bar W = 21.75$ its mean, and $\epsilon$ is the residual. The eleven monthly offsets are equivalent to a complete Fourier series of the annual cycle, and so absorb any recurring pattern with a period of one year or a fraction of it, including semester and term structure. In the category fits of Sec.~\ref{sec:fields_method}, truncated Fourier series with one to three harmonics are disfavoured relative to the monthly offsets by $\Delta\mathrm{BIC} = +6$ to $+37$ in every group tested except statistics. The weekday term captures the dependence of a month's count on how weekends fall in a given year, which varies between $20$ and $23$ weekdays and is not periodic in the calendar month. It lowers the BIC of the final aggregate model by $65$, and the fitted elasticity is $w = 0.87$ for the aggregate and $0.2$--$0.8$ for the larger category groups. The growth rate in each segment is $\exp(b + \sum_{\tau_j < t} c_j) - 1$ per year. The model is fitted by ordinary least squares to data from January 2010 onwards, a period in which the category composition is broadly stable. Two knots fixed at April 2020 and April 2021 absorb the pandemic excess and subsequent correction.

\subsection{Break detection}
\label{sec:breaks_method}

Additional knots are located by exhaustive search on a monthly grid. For each candidate configuration we compute the Bayesian information criterion, $\mathrm{BIC} = n \ln(\mathrm{RSS}/n) + p \ln n$, where $n$ is the number of months and $p$ the number of parameters. Configurations are compared by $\Delta\mathrm{BIC}$, and the uncertainty in each knot location is quoted as the range over which $\Delta\mathrm{BIC} < 2$ relative to the optimum. The residuals of the final aggregate model have lag-one autocorrelation $\rho = 0.21$. We therefore also recompute each $\Delta\mathrm{BIC}$ with the effective number of independent months, $n_\mathrm{eff} = n(1-\rho)/(1+\rho)$, and quote these values where they bear on the conclusions.

\subsection{Field comparison}
\label{sec:fields_method}

For each category group we fit Eq.~\eqref{eq:model} without knots to the interval January 2022 to November 2025 and compute the mean residual over December 2025 to September 2026, which we refer to as the excess. Its significance is quoted as $z = \bar\epsilon/(\sigma/\sqrt{10})$, with $\sigma$ the residual scatter in the fitting interval. We also compute the change in year-on-year growth of January--September totals between 2025 and 2026, which does not depend on the trend model. Finally, a single additional knot is located in each group by the procedure of Sec.~\ref{sec:breaks_method}, searching from January 2023 to June 2026. The difference in excess between the no-CS-listing group and the CS group provides a difference-in-differences estimate of the effect outside AI research, with CS acting as the group whose growth is dominated by its own research boom. We also test whether the field ordering of the excess follows survey-based estimates of agentic-tool adoption~\cite{RepublicOfScience2026}.

\subsection{Author concentration}
\label{sec:authors_method}
Using the full metadata for mathematics in September 2025 and September 2026, we count the papers on which each distinct author name appears in the month. Names are normalised by removing diacritics, punctuation and case. Two effects complicate comparison between months of different volume: an individual author's count rises if their output rises, which is the signal of interest, but the number of authors exceeding any fixed threshold also rises with sample size, and distinct individuals sharing a name are merged more often in a larger sample. We therefore report each statistic for the full samples and for random subsamples of September 2026 matched in size to September 2025, averaged over twenty realisations. Growth in the number of papers $P$ is decomposed exactly as
\begin{equation}
\ln\frac{P_{26}}{P_{25}} = \ln\frac{A_{26}}{A_{25}} + \ln\frac{\bar n_{26}}{\bar n_{25}} - \ln\frac{\bar a_{26}}{\bar a_{25}},
\label{eq:decomp}
\end{equation}
where $A$ is the number of distinct authors, $\bar n$ the mean number of papers per author, and $\bar a$ the mean number of authors per paper.

\subsection{Author-count series}
\label{sec:solo_method}
To follow authorship over time we sampled the author lists of papers in four groups, all without a CS listing, for every month from January 2022 to September 2026. Mathematics and condensed matter have large submission excesses; quantitative biology is the one group whose output fell; and experimental high-energy physics (\texttt{hep-ex}), which shows no 2025--2026 break in submissions, serves as a control. For each month the API was queried in order of submission date at four evenly spaced offsets of $250$ papers each, so that each sample spans the whole month. This gives $1\,000$ papers per month for mathematics and condensed matter and every paper in the month for the two smaller groups (on average $261$ and $197$). For mathematics the sampled values for September 2025 and September 2026, $2.35$ and $2.06$ authors per paper, agree with the full metadata of Sec.~\ref{sec:authors_method} ($2.33$ and $2.09$) within sampling error. We use the fraction of solo-authored papers rather than the mean number of authors per paper, which in condensed matter and \texttt{hep-ex} is dominated by large collaborations. Breaks are located as in Sec.~\ref{sec:breaks_method} on a linear trend, allowing either a change of slope or a change of level, and the excess is computed as in Sec.~\ref{sec:fields_method}. Multiplying the solo fraction by each group's monthly count gives the numbers of solo-authored and multi-author papers.

\section{Results}

\begin{figure*}[t]
\includegraphics[width=\textwidth]{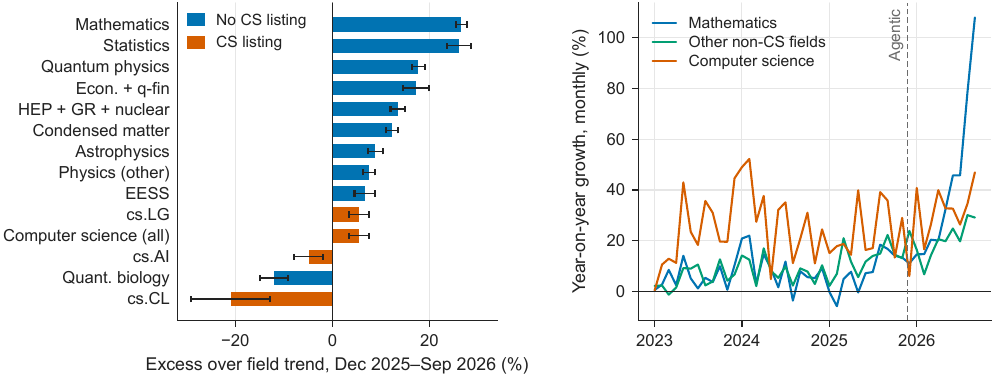}
\caption{Left: mean excess of monthly submissions over each group's January 2022--November 2025 trend, averaged over December 2025--September 2026, with $\pm 1\sigma$ uncertainty $\sigma/\sqrt{10}$. Groups without a CS listing are shown in blue, CS groups in vermilion. Right: monthly year-on-year growth for mathematics without a CS listing, all other non-CS groups combined, and computer science. The dashed line marks the cluster of agentic model releases in late 2025.}
\label{fig:fields}
\end{figure*}

\subsection{Aggregate series}

Figure~\ref{fig:aggregate} shows the aggregate series together with the extrapolation of a single exponential plus seasonal model fitted to January 2010 to November 2022. Over the fitting interval the growth rate is $9.0\%$ per year with residual scatter $\sigma = 0.049$ in $\ln N$. The data fall $2$--$5\%$ below the extrapolation in 2023, the year following the pandemic correction, and above it thereafter by an amount that increases each year, reaching $9$--$12\%$ in 2025 and $28$--$34\%$ over January to September 2026, depending on the start of the fitting window (2005--2015). September 2026 lies $10\sigma$ above the extrapolation.

The year-on-year growth of January--September totals was $11.2\%$ in 2023, $18.2\%$ in 2024, $16.6\%$ in 2025 and $30.0\%$ in 2026. Over 1997--2022 annual growth had mean $10.0\%$ and standard deviation $5.0\%$, so the 2024 and 2025 values are individually unremarkable while the 2026 value lies $4\sigma$ from the historical mean.

\begin{table*}
\caption{Field comparison. Trend: growth rate fitted to January 2022--November 2025. Excess: mean residual over December 2025--September 2026 with significance $z$. YoY: growth of January--September totals relative to the previous year. All non-CS groups exclude papers with any \texttt{cs.*} listing; HEP subgroups are shown individually. Break locations for each group are listed in Table~\ref{tab:breaks}.}
\label{tab:results}
\begin{ruledtabular}
\begin{tabular}{lrrrrr}
Group & Trend & Excess & $z$ & YoY 2025 & YoY 2026 \\
 & (\%/yr) & (\%) & & (\%) & (\%) \\
\hline
All & 15.3 & 12.1 & 8.9 & 16.9 & 30.4 \\
Computer science & 24.2 & 5.5 & 2.6 & 23.9 & 33.0 \\
\quad cs.LG & 17.7 & 5.5 & 2.6 & 20.0 & 25.3 \\
\quad cs.AI & 46.6 & $-4.9$ & $-1.7$ & 40.5 & 42.7 \\
\quad cs.CL & 44.1 & $-21.0$ & $-2.9$ & 21.0 & 19.2 \\
No CS listing & 8.1 & 17.1 & 16.2 & 10.4 & 27.6 \\
\quad Mathematics & 7.3 & 26.5 & 21.0 & 6.1 & 43.6 \\
\quad Statistics & 10.0 & 26.1 & 9.6 & 15.1 & 38.3 \\
\quad Quantum physics & 12.3 & 17.7 & 12.2 & 13.3 & 33.6 \\
\quad Economics & 15.9 & 17.2 & 6.0 & 22.5 & 33.9 \\
\quad HEP, GR, nuclear & 4.4 & 13.5 & 8.2 & 8.3 & 18.0 \\
\qquad hep-th & 4.6 & 14.5 & 8.3 & 4.3 & 21.6 \\
\qquad gr-qc & 6.7 & 11.8 & 6.2 & 8.8 & 19.5 \\
\qquad hep-ex & 1.7 & 14.6 & 3.4 & 13.8 & 10.6 \\
\quad Condensed matter & 5.7 & 12.3 & 9.9 & 9.2 & 17.1 \\
\quad Astrophysics & 7.5 & 8.8 & 5.3 & 12.1 & 14.3 \\
\quad Other physics & 7.1 & 7.5 & 5.9 & 10.4 & 12.6 \\
\quad EESS & 18.8 & 6.6 & 3.1 & 22.9 & 25.3 \\
\quad Quant.\ biology & 13.9 & $-12.0$ & $-4.4$ & 19.7 & $-3.4$ \\
\end{tabular}
\end{ruledtabular}
\end{table*}

\subsection{Breaks in the growth rate}

Allowing one post-pandemic knot reduces the BIC by $83$ relative to the model with pandemic knots alone ($52$ with $n_\mathrm{eff}$). Its preferred location is the turn of 2022--2023, with a $\Delta\mathrm{BIC}<2$ interval from October 2022 to February 2023, and the growth rate after it is approximately $16\%$ per year. A two-knot search reduces the BIC by a further $41$ ($25$ with $n_\mathrm{eff}$). In this configuration the first knot moves to mid-2022, before the release of ChatGPT~\cite{OpenAIChatGPT2022} and coincident with the end of the post-pandemic trough, while the second is located between December 2025 and February 2026. Excluding September 2026 leaves the second knot in the same interval, with $\Delta\mathrm{BIC} = -20$ ($-12$ with $n_\mathrm{eff}$). We therefore regard the second break as a robust feature of the data, while the first cannot be separated from the recovery from the pandemic correction.

The second break follows a cluster of releases of models designed for extended autonomous work in the final quarter of 2025~\cite{AnthropicOpus45}. Earlier capable models released in mid-2024 and mid-2025 are not associated with detectable breaks in the aggregate: with the two knots fixed, adding a third anywhere between January 2023 and November 2025 increases the BIC by at least $4$. If the association is causal, the response therefore has a threshold character, appearing when tools became able to execute multi-step research and writing tasks rather than when conversational assistance became available.

\subsection{Field dependence}

Table~\ref{tab:results} and Fig.~\ref{fig:fields} summarise the field comparison. The central result is that the two aggregate breaks have different origins.

Papers with no CS listing grew at $8.1\%$ per year between January 2022 and November 2025, consistent with the historical rate of the archive as a whole. Over the same interval papers with a CS listing grew at $24\%$ per year, cs.AI at $47\%$ and cs.CL at $44\%$. The acceleration of the aggregate after 2022 is therefore attributable entirely to growth in the volume of computer-science and AI research; the period of widespread conversational LLM use left no detectable trace in the volume of submissions in other fields.

After November 2025 the pattern reverses. Papers with no CS listing lie $17.1\%$ above their trend ($z = 16.2$), and their January--September year-on-year growth rises from $10.4\%$ in 2025 to $27.6\%$ in 2026. A single-knot search places the break in February 2026 with $\Delta\mathrm{BIC} = -92$. The CS group shows a smaller excess of $5.5\%$ ($z = 2.6$) and a weaker break ($\Delta\mathrm{BIC} = -8$, located between November 2025 and June 2026), while cs.AI and cs.CL show no excess and no 2026 break. The difference-in-differences estimate of the excess outside CS relative to CS is $+11.6$ percentage points; in terms of acceleration of year-on-year growth it is $+8.2$ percentage points. The 2026 surge is therefore not a continuation of the AI-research boom; it is an increase in the rate of production of papers in other fields.

Within the no-CS-listing group the response is strongly field dependent. Mathematics has the largest excess ($27\%$, $z = 21$) and the sharpest and latest break, in March 2026 ($\Delta\mathrm{BIC}<2$ interval March--April 2026, $\Delta\mathrm{BIC} = -123$). Its monthly year-on-year growth, which had been near $15\%$ through February 2026, rises to $32\%$ in May, $46\%$ in June and July, $79\%$ in August and $108\%$ in September. Mathematics accounts for $13\,726$ of the $28\,041$ additional non-CS submissions between January--September 2025 and January--September 2026, and for $62\%$ of the non-CS increase between September 2025 and September 2026. Statistics ($26\%$), quantum physics ($18\%$), economics ($17\%$), high-energy physics, gravitation and nuclear physics ($14\%$) and condensed matter ($12\%$) follow, while astrophysics ($9\%$), other physics ($7\%$) and electrical engineering and systems science ($7\%$) show smaller excesses and quantitative biology a deficit. Breaks in the physics groups fall between September 2025 and February 2026; in statistics, economics, astrophysics and EESS the single-knot search instead identifies an earlier, gradual steepening between late 2024 and mid-2025.

The ordering does not follow measured rates of adoption of agentic tools. In the one available survey~\cite{RepublicOfScience2026}, economists and social scientists had the highest adoption of a widely used agentic coding tool by February 2026 and mathematicians the lowest. Mathematics nevertheless shows the largest response. The ordering instead follows the extent to which research in a field consists of formal reasoning and writing rather than experiment or observation. A direct test within a single community is provided by high-energy physics, where theory (\texttt{hep-th}) and experiment (\texttt{hep-ex}) share conferences, funding and publication culture. Excluding CS listings, the January--September year-on-year growth of \texttt{hep-th} rises from $4.3\%$ in 2025 to $21.6\%$ in 2026, with a break located in December 2025 ($\Delta\mathrm{BIC}<2$ interval November 2025--January 2026, $\Delta\mathrm{BIC}=-42$). For \texttt{hep-ex} the same quantity falls from $13.8\%$ to $10.6\%$, and the single-knot search places its only break in early 2023, with none in 2025--2026. The mean residual of \texttt{hep-ex} over December 2025--September 2026 is nonetheless $+15\%$, comparable to that of \texttt{hep-th} ($+14\%$), but it is measured against a nearly flat and noisy fitted trend ($1.7\%$ per year) and is of low significance ($z=3.4$ against $8.3$). On the two measures that do not depend on the fitted trend, the split therefore supports the gradient; on the trend residual it is inconclusive. \texttt{gr-qc}, which is predominantly theoretical, shows a similar acceleration in year-on-year growth, from $8.8\%$ to $19.5\%$, although its single break is located earlier, in the first half of 2025. Within mathematics the same gradient is visible across subfields: between September 2025 and September 2026 submissions with primary classification in differential geometry increased by a factor $2.9$, functional analysis, number theory, complex variables and classical analysis by $2.4$--$2.5$, combinatorics by $2.3$, and algebraic geometry by $2.0$, while numerical analysis and optimisation and control, the most computational subfields, increased by $1.6$.

\subsection{Concentration among authors}
\label{sec:authors}

The arXiv announcement accompanying the per-submitter limit attributes the surge in part to a small number of authors submitting very large numbers of papers~\cite{arXivCap2026}. For mathematics, the field with the largest increase, the metadata do not support this as the principal mechanism.

Between September 2025 and September 2026 the number of mathematics papers rose from $3\,945$ to $8\,203$ and the number of distinct authors from $8\,267$ to $13\,536$. The mean number of papers per author rose from $1.11$ to $1.27$ and the mean number of authors per paper fell from $2.33$ to $2.09$. By Eq.~\eqref{eq:decomp}, $67\%$ of the logarithmic growth is due to the larger number of distinct authors, $18\%$ to higher output per author and $15\%$ to smaller author teams. Solo-authored papers increased by a factor $2.7$, from $1\,142$ ($29\%$) to $3\,049$ ($37\%$).

Output per author is nonetheless more concentrated than before. In the full samples, the number of authors appearing on more than five papers in the month rose from $3$ to $53$, the largest individual count from $6$ to $12$, and the fraction of papers with at least one author appearing on three or more papers from $7.8\%$ to $23.5\%$. Part of this rise is a consequence of the doubling of the sample. In size-matched subsamples of September 2026 the corresponding fraction is $10.8\%$ and the number of authors with more than five papers is $8$, compared with $7.8\%$ and $3$ in September 2025. Concentration has increased, but the papers of high-output authors are a minority of the excess, and the distribution of the increase over the fourteen largest subfields, with ratios between $1.6$ and $2.9$, argues against a small number of sources.

\begin{figure}[t]
\includegraphics[width=\columnwidth]{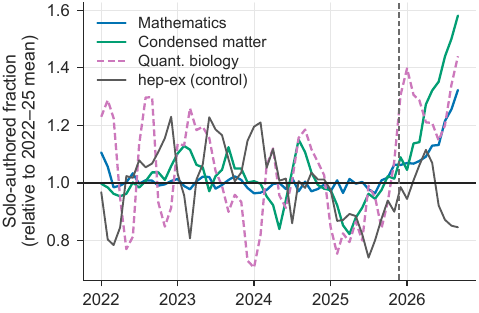}
\caption{Fraction of solo-authored papers, relative to each group's January 2022--November 2025 mean, as a three-month centred average. All groups exclude papers with a CS listing. Mathematics and condensed matter are samples of $1\,000$ papers per month; quantitative biology and \texttt{hep-ex} include every paper. The dashed line marks the cluster of agentic model releases in late 2025.}
\label{fig:authors}
\end{figure}

\begin{table*}[t]
\caption{Single authorship by group. Solo fraction: mean share of solo-authored papers over January 2022--November 2025, over December 2025--September 2026, and in September 2026, with significance $z$ of the excess. Break: best single break in the solo fraction, with $\Delta\mathrm{BIC}$. YoY: growth of January--September counts of solo-authored and multi-author papers relative to the previous year. All groups exclude papers with a CS listing.}
\label{tab:solo}
\begin{ruledtabular}
\begin{tabular}{lrrrrlrrrr}
 & \multicolumn{3}{c}{Solo fraction (\%)} & & & \multicolumn{2}{c}{Solo YoY (\%)} & \multicolumn{2}{c}{Multi-author YoY (\%)} \\
Group & 2022--25 & Dec--Sep & Sep 2026 & $z$ & Break & 2025 & 2026 & 2025 & 2026 \\
\hline
Mathematics & 28.8 & 32.9 & 37.7 & 9.0 & Sep 2025 ($-37$) & +8 & +70 & +6 & +33 \\
Condensed matter & 7.6 & 9.8 & 11.8 & 9.7 & Dec 2025 ($-44$) & +0 & +71 & +10 & +13 \\
Quant.\ biology & 11.7 & 15.1 & 16.9 & 4.8 & Nov 2025 ($-9$) & +5 & +44 & +22 & $-9$ \\
hep-ex (control) & 23.6 & 22.8 & 19.9 & 0.7 & none in 2025--26 & $-11$ & +28 & +23 & +6 \\
\end{tabular}
\end{ruledtabular}
\end{table*}

\subsection{A shift to single-author papers}
\label{sec:solo}

The decomposition of Sec.~\ref{sec:authors} shows that the share of solo-authored papers in mathematics rose between the two Septembers. The monthly series of Sec.~\ref{sec:solo_method} show that this is a break rather than a drift, and that it is not confined to mathematics (Fig.~\ref{fig:authors} and Table~\ref{tab:solo}).

In mathematics the solo-authored fraction was flat at $28.8\%$ from January 2022 to November 2025 and has risen since, reaching $37.7\%$ in September 2026. The best single break is a change of slope in September 2025 ($\Delta\mathrm{BIC}<2$ interval July--November 2025, $\Delta\mathrm{BIC} = -37$), about six months before the break in mathematics submissions, and the mean excess over December 2025--September 2026 is $+4.6$ percentage points ($z = 9.0$). Condensed matter, where solo authorship is much rarer, shows the same pattern: the fraction rose from $7.6\%$ to $11.8\%$ in September 2026, with a break in December 2025 ($\Delta\mathrm{BIC} = -44$, $z = 9.7$) that coincides with the break in its submissions.

Counting papers makes the change clearer. Between January--September 2025 and the same months of 2026, solo-authored papers increased by $70\%$ in mathematics and $71\%$ in condensed matter, against $8\%$ and $0\%$ the year before, while multi-author papers increased by $33\%$ and $13\%$. Solo-authored papers were $28.5\%$ of mathematics output in 2025 but account for $46\%$ of its increase; in condensed matter they were $7\%$ of output and account for $29\%$ of the increase.

Quantitative biology is the most informative case. Its total output fell by $3.4\%$ between the two periods, the only deficit in Table~\ref{tab:results}, yet its solo-authored papers increased by $44\%$, against $5\%$ the year before. The fall is entirely in multi-author papers, which declined by $9\%$ after growing by $22\%$ in 2025, and the solo fraction shows a level shift in November--December 2025 ($\Delta\mathrm{BIC} = -9$, $z = 4.8$). An increase in single-author output is therefore present even in a field whose overall output contracted, and the aggregate count conceals it. We have not tested explanations for the fall in multi-author papers in this field.

\texttt{hep-ex}, the control, shows no such change. Its solo-authored fraction over December 2025--September 2026 is consistent with its earlier mean ($z = 0.7$), the single-break search finds no break in 2025--2026, and over the two years to January--September 2026 its solo-authored papers grew by $13\%$ against $30\%$ for multi-author papers. The shift towards single authorship therefore appears in fields with very different team sizes and research practices, coincides in time with the break in submissions, and is absent from experimental particle physics, where papers report measurements made by collaborations.

\subsection{The per-submitter limit}
\label{sec:cap}
From 1 October 2026 each submitter may make at most two new submissions per calendar month, with no more than three active at any time; rejected submissions count towards the monthly limit~\cite{arXivCap2026}. The limit binds on the submitting account rather than on authors, so the public metadata cannot identify affected papers directly. They do provide an upper bound: a paper can only have been displaced by the limit if at least one of its authors appeared on three or more papers in the month. In September 2026 this was the case for $23.5\%$ of mathematics papers, and for $10.8\%$ in a sample of the size of September 2025. If, as the decomposition of Sec.~\ref{sec:authors} suggests, most of the excess arises from a larger population of authors each producing one or two papers, the limit should reduce the volume only modestly. The early October data are consistent with this. The published count for October 2026 stood at $31\,672$ on 23 October, already $14\%$ above the total for the whole of October 2025 ($27\,692$) with more than a week of the month remaining. Per weekday, submissions in October to date have run at or above the September rate, so the limit has produced no visible reduction in volume in its first three weeks. The complete October count, and the excess relative to the extrapolated trend by field, will be reported in a revised version of this paper.


\section{Discussion}

\subsection{Two phenomena}

The field comparison separates two phenomena that are conflated in the aggregate record. From 2023 to 2025 arXiv grew faster than its historical rate because research on AI itself expanded; the introduction of conversational LLMs, whatever its effect on the prose of individual papers, did not measurably change the number of papers produced in other fields. From late 2025 the number of papers in other fields began to rise sharply, while the AI categories, which had driven the earlier acceleration, did not. The second phenomenon is the one for which arXiv has introduced a submission limit.

\subsection{What the pattern implies}

Three features of the second phenomenon are informative about its cause. Its timing follows, by two to four months, the release in late 2025 of models designed for extended autonomous work, and no earlier release is associated with a comparable break. Its magnitude is ordered by the extent to which a field's research output consists of formal reasoning and text: largest in mathematics and statistics, intermediate in theoretical physics, smallest in experimental and observational fields, and largest within mathematics for the least computational subfields. And it is carried by a larger number of active authors and, across fields, by a shift to single authorship that is absent from experimental particle physics, rather than by a few prolific sources. These features are what one would expect if the change in 2025 was in the ability of available tools to carry a research problem, rather than only its write-up, to completion, and if the effect is largest where the research product can be produced entirely at a desk. They are not what one would expect from an expansion of a single research area, from changes in arXiv's moderation, or from bulk submission by a small number of accounts. That the ordering does not follow surveyed adoption of agentic coding tools~\cite{RepublicOfScience2026} suggests that the relevant variable is the capability of the tools in a field's core task rather than the uptake of a particular product.

\subsection{Limitations}

The design has corresponding limitations. Because it is agnostic of LLM detection it measures the aggregate response of submission volume to the availability of agentic tools, and cannot identify which papers were produced with them or assess their quality. The attribution rests on timing and on the pattern across fields; an unrelated change affecting formal fields from early 2026 would produce a similar signature, although we are not aware of a candidate. Category counts include cross-lists, and the public API does not identify the submitting account, so the author-level analysis is a proxy for the submitter-level behaviour on which the arXiv limit acts. Finally, the advantage of calendar timing over detection-triggered timing~\cite{Renault2026} is obtained at the price of a single treatment date for all authors, so the estimates here are of the population-level response, not of the effect on an individual adopter. The author-count series are samples for two of the four groups, and the solo-authored fraction is a coarse measure of how work is divided within a paper.

\subsection{A data point}

The production of this paper provides a single, anecdotal but concrete measurement of the cost of agentic research. The interval from the first question posed to the model, at 14:36 on 22 October 2026, to a complete draft including data retrieval from the arXiv API, all statistical analyses and figures, and the text, was about 3.75 hours of wall-clock time. About 80 minutes of this was rate-limited data retrieval, which ran in parallel with the drafting. The model made 176 tool invocations, which together issued about 1\,100 count queries to the API and retrieved some 12\,000 metadata records, while the author's own input consisted of some 60 interactions with the model. A comparable analysis carried out conventionally would, in the author's estimate, have occupied at least a working week. A reduction of this order in the time cost of a paper of this type is sufficient to account for the per-author increases in output reported in Refs.~\cite{Kusumegi2025,arXivCap2026}, and illustrates why a limit on submissions per submitter, rather than a limit on the effort required per submission, has become the binding constraint.

\section{Open questions}
\label{sec:open}

The submission record raises wider questions that it cannot answer. The first concerns who does the work. Single-author papers have risen sharply since late 2025 in mathematics, condensed matter and quantitative biology (Sec.~\ref{sec:solo}), and in mathematics the mean number of authors per paper fell from $2.33$ to $2.09$ in a year. Submission metadata do not record career stage, so we cannot tell whether established researchers are now completing work that would previously have involved students and postdoctoral researchers. If they are, the effect on the training of the next generation would be slow to appear in publication counts and difficult to reverse. Related signals exist outside arXiv. Employment of workers aged 22--25 in the occupations most exposed to AI has fallen by $16\%$ relative to that of older workers in the same occupations~\cite{Brynjolfsson2025}. The fraction of US STEM doctoral dissertations whose abstracts contain AI-generated text rose from $1.7\%$ in 2023 to $29\%$ in the first five months of 2026, and graduates whose dissertations are affected are less likely to remain in academia~\cite{Gross2026}. Doctoral admissions at major US research universities fell by $15\%$ for autumn 2026~\cite{AAU2026}. That decline is attributed principally to reductions in federal research funding, and whether AI adoption contributes to it has not been examined.

The second question is whether the present system can absorb the increase. If the doubling seen in mathematics extends to most fields, peer review, which depends on unpaid effort drawn from the same researchers whose output is growing, is unlikely to keep pace. Automated review is the obvious response, and it is already in use at scale: an estimated $21\%$ of the reviews submitted to a major machine-learning conference in late 2025 were generated entirely by AI~\cite{NatureICLR2025}. A literature written and assessed largely by machines raises the question of how results will be recorded, disseminated and attributed; how the credit on which hiring, promotion and funding depend will be assigned; and how researchers will survey a literature whose volume exceeds what any person can read.

Attribution also has a legal dimension. Under US law, material generated by an AI system without sufficient human creative control is not eligible for copyright~\cite{USCO2025,Thaler2025}. As the human contribution to a paper shifts from execution towards direction, the boundary between authored and generated work, and with it the basis of individual intellectual credit, will become harder to draw. The outcome may be a renegotiated model of individual authorship, or a move towards collective credit for results produced with shared tools; either would change what an academic career consists of. These questions concern the purpose of academic research as much as its mechanics. The pace of the change in the submission record suggests that they will need answers within years rather than decades.

\section{Conclusions}

The arXiv submission record, analysed without any attempt to detect machine-generated text, contains a distinct break in early 2026 that is separate from the growth of AI research, follows the arrival of agentic models, is concentrated in fields whose research can be carried out entirely at a desk, and is accompanied by a shift to single-author papers. Three consequences follow. For arXiv, a limit on submissions per account addresses concentration among prolific submitters, but the mathematics data suggest that most of the increase arises from a broad population of authors each submitting one or two papers, which such a limit will not reduce; moderation capacity will need to scale with volume regardless. For the evaluation of researchers, publication counts in the affected fields after late 2025 are no longer comparable with those before it, and assessments that rely on them will need to account for the change. For mathematics in particular, the doubling of submissions within a year is an unprecedented change in the output of a field whose conventions of proof and review are among the most demanding in science; whether the additional papers represent a corresponding increase in established results is a question the submission record alone cannot answer, and one that the mathematical community is best placed to examine. The wider questions raised by the change, concerning the training of researchers, the capacity of peer review and the basis of scholarly credit, are set out in Sec.~\ref{sec:open}.

\section*{Data availability}

The monthly category counts used in this paper, the September 2025 and September 2026 mathematics metadata summaries, the sampled monthly author counts, and the analysis code are available from the author on request and will be deposited in a public repository on publication. All underlying data are publicly available through the arXiv API.

\begin{acknowledgments}
We thank arXiv for use of its open interoperability.
Data retrieval, statistical analysis, figure preparation and the first draft of the text were carried out with Claude Opus 5.5 (Anthropic), operating as an agent with access to a code-execution environment, a web browser and web search, under the direction of the author, who specified the questions, reviewed the analysis and is responsible for the content.
\end{acknowledgments}

\appendix

\section{Break locations by group}
\label{app:breaks}

Table~\ref{tab:breaks} lists, for each category group, the single additional break located by the procedure of Sec.~\ref{sec:breaks_method} in the January 2022--September 2026 series, searching from January 2023 to June 2026, together with the fitted weekday elasticity $w$ of Eq.~\eqref{eq:model}.

\begin{table}[h]
\caption{Single additional break located in each group's January 2022--September 2026 series: best location, $\Delta\mathrm{BIC}<2$ interval, and $\Delta\mathrm{BIC}$ relative to the model without a break. $w$: fitted weekday elasticity, a nuisance parameter that absorbs calendar variation; its uncertainty is $\pm 0.13$--$0.20$ for the larger groups and $\pm 0.3$--$0.5$ for the smaller ones, so differences between groups are not significant. All non-CS groups exclude papers with any \texttt{cs.*} listing.}
\label{tab:breaks}
\begin{ruledtabular}
\footnotesize
\begin{tabular}{@{}llllr@{}}
Group & Break & Interval & $\Delta\mathrm{BIC}$ & $w$ \\
\hline
All & Feb 2026 & Jan--Mar 2026 & $-51$ & 0.33 \\
Computer science & Mar 2026 & Nov 2025--Jun 2026 & $-8$ & 0.18 \\
\quad cs.LG & Jun 2026 & Dec 2025--Jun 2026 & $-8$ & 0.29 \\
\quad cs.AI & Feb 2024 & Sep 2023--Jan 2025 & $-0.4$ & 0.11 \\
\quad cs.CL & Feb 2024 & Nov 2023--Aug 2024 & $-10$ & 0.60 \\
No CS listing & Feb 2026 & Feb 2026 & $-92$ & 0.49 \\
\quad Mathematics & Mar 2026 & Mar--Apr 2026 & $-123$ & 0.52 \\
\quad Statistics & May 2025 & Dec 2024--Aug 2025 & $-49$ & 0.70 \\
\quad Quantum physics & Sep 2025 & Aug--Nov 2025 & $-61$ & 0.55 \\
\quad Economics & Dec 2024 & Aug 2024--Feb 2025 & $-34$ & $-0.01$ \\
\quad HEP, GR, nuclear & Dec 2025 & Nov 2025--Jan 2026 & $-43$ & 0.47 \\
\qquad hep-th & Dec 2025 & Nov 2025--Jan 2026 & $-42$ & 0.30 \\
\qquad gr-qc & Mar 2025 & Dec 2024--Jun 2025 & $-31$ & 0.13 \\
\qquad hep-ex & Mar 2023 & Feb--May 2023 & $-22$ & 1.08 \\
\quad Condensed matter & Nov 2025 & Sep 2025--Feb 2026 & $-38$ & 0.37 \\
\quad Astrophysics & Sep 2024 & Jul--Dec 2024 & $-29$ & 0.82 \\
\quad Other physics & Feb 2026 & Jan--Mar 2026 & $-18$ & 0.70 \\
\quad EESS & Oct 2024 & Apr 2024--Jan 2025 & $-11$ & $-0.02$ \\
\quad Quant.\ biology & Sep 2025 & Apr--Dec 2025 & $-4$ & $-0.16$ \\
\end{tabular}
\end{ruledtabular}
\end{table}

\bibliography{refs}

\end{document}